\documentclass[11pt]{article}

\columnwidth\textwidth
\usepackage{graphicx}
\usepackage{epsfig}
\usepackage{amsmath}
\usepackage{amssymb}
\usepackage{times}
\usepackage{dsfont}

\begin{document}

\title{Focusing of high-energy particles\\
in the electrostatic field of a homogeneously charged sphere\\
and the effective momentum approximation}
\author{Andreas Aste, Dirk Trautmann\\
Department of Physics and Astronomy, University of Basel,\\
Klingelbergstrasse 82, 4056 Basel, Switzerland}
\date{July 4, 2007}

\maketitle

\begin{center}
\begin{abstract}
The impact of the strongly attractive electromagnetic field of heavy
nuclei on electrons in quasi-elastic $(e,e')$ scattering is often accounted for by
the effective momentum approximation.
This method is a plane wave Born approximation which
takes the twofold effect of the attractive nucleus on initial and final state
electrons into account, namely the modification of the
electron momentum in the vicinity of the nucleus, and the focusing of electrons
towards the nuclear region leading to an enhancement of the corresponding
wave function amplitudes.
The focusing effect due to the attractive Coulomb field of a homogeneously
charged sphere on a classical ensemble of charged
particles incident on the field is calculated in the highly relativistic
limit and compared to results obtained from exact solutions
of the Dirac equation. The result is relevant for the theoretical
foundation of the effective momentum approximation and describes the
high energy behavior of the amplitude of continuum Dirac waves in the potential
of a homogeneously charged sphere. Our findings indicate that
the effective momentum approximation is a useful approximation for the calculation
of Coulomb corrections in $(e,e')$ scattering off heavy nuclei for sufficiently
high electron energies and momentum transfer.

\end{abstract}
\end{center}

\noindent 

\section{Introduction}
Scattering experiments can be viewed as one of the very important tools of experimental
particle physics since the famous Lord Ernest Rutherford scattering experiment
of $\alpha$-particles off the nuclei within a gold foil in 1911 \cite{Rutherford}.
To explore the structure of the nucleus, the main tool used today is
electron scattering due to the transparency of the nuclear volume for electrons.
E.g., inclusive $(e,e')$ scattering, where only the final
electron is observed, provides information about
the nuclear Fermi momentum by measuring the width of the quasi-elastic peak \cite{Whitney},
or the high-momentum components of nucleon wave functions when the tail
of the quasi-elastic peak is investigated \cite{Benhar95,Rohe}.
Information about infinite nuclear matter is obtained
by extrapolating the mass number $A \rightarrow \infty$ \cite{Day89}, and
possible modifications of the nucleon form factors inside a nucleus
are related to the Coulomb sum rules \cite{MezianiCoulomb}.
However, although electrons with energies of typically some hundred MeV
are used in the experiments, the distortion of the electron wave functions
due to the strongly attractive electrostatic field of heavy nuclei can no
longer be neglected, such that calculations in the plane wave Born
approximation (PWBA) are no longer reliable.

Calculations using exact Dirac wave functions are feasible but cumbersome
and difficult compared to the PWBA calculations. As a consequence,
various approximate methods have been proposed in the past
for the treatment of Coulomb distortions
\cite{Lenz,Knoll,Giusti,Rosenfelder,Giusti88,Traini88,Traini95,
Rosenfelder80,Kosik},
and there is an extensive literature on the so-called eikonal approximation
\cite{Yennie64,Sucher,Blankenbecler,Wallace1,Wallace2,Abarbanel,Aste0,Aste1,Aste2}.

In this paper, we give a concise classical derivation of the effective momentum
approximation (EMA), which has the advantage that one works with plane waves and 
which plays an important role in experimental data analysis. The classical
high-energy results are compared to results obtained from exact solutions
of the Dirac equation. Our findings concerning the correct use of the EMA
are of actual importance, since  there is now considerable theoretical and
experimental interest  in extracting longitudinal and transverse structure
functions as a function of energy loss for fixed three-momentum transfer for
a range of nuclei. 
Recently, a Thomas Jefferson National Accelerator Facility (TJNAF) proposal
for quasi-elastic electron scattering measurement in the momentum transfer range
$0.55$ GeV/c $\le$ $|\vec{q}|$ $\le$ $1.0$ GeV/c was approved
such that the experiments will be performed in the near future
using $^4$He, $^{12}$C, $^{56}$Fe and $^{208}$Pb as target nuclei \cite{jlab}.

We shortly comment qualitatively on the connection
between the distorted wave Born approximation (DWBA) and the EMA and its correct
application. In DWBA, one calculates matrix elements with exact initial and final
state electron wave functions. Unlike the plane waves with constant amplitude
used in the PWBA, these wave functions are focused towards the nuclear
region, and the local electron momenta are enhanced there due to the attractive
positively charged nucleus. In the EMA, the focusing and the momentum transfer
in the relevant nuclear region, where the nucleons get knocked, are accounted for
by effective (average) values. It is important to mention that one must base
EMA calculations indeed on average values, although in the literature, the use of
effective values for the focusing and the effective momenta valid
only in the center of the nucleus is widespread. However, the choice of
such values is not appropriate, as will be explained in detail in this paper.

There are two equivalent methods for the correct application of the EMA.
First, one may calculate the cross section from the corresponding
theoretical PWBA expression for the $(e,e')$ scattering cross section
with effective momenta. This introduces an
artificially enhanced phase space for the final state electron,
since also in the DWBA, the phase space is given by the undistorted
asymptotic momenta of the final state particles. However,
this enhanced phase space accidentally accounts for the focusing effect
on the final state electron with a high level of accuracy.
Because the initial focusing has not yet been taken into account, one has to multiply
the cross section calculated so far additionally by the effective focusing factor
of the initial state electron.

Another equivalent approach is to factorize the theoretical expression for
the $(e,e')$ cross section into the Mott cross section given by eq. (\ref{Mott2})
and a response function according to eq. (\ref{Mott1}).
The interesting point is that the impact of the focusing cancels against
the modification of the momentum transfer in the Mott cross section.
Accordingly, one may calculate the EMA cross section by leaving the Mott part
unchanged and by evaluating the response function with the momenta replaced
by their effective values.

A critical overwiew on the history of the effective momentum approximation
and its correct and incorrect application can be found in \cite{Traini2001}.

\section{Quasi-elastic scattering}
\begin{figure}[htb]
        \centering
        \includegraphics[width=10.0 cm]{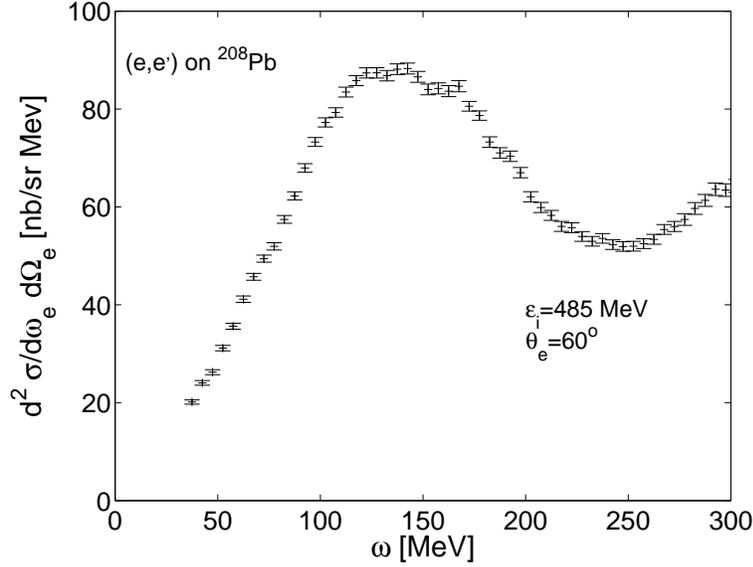}
        \caption{Quasi-elastic $(e,e')$ scattering cross section data taken at
                 Saclay for initial electron energy $\epsilon_i=485$ MeV
                 and electron scattering angle $\Theta_e=60^o$.}
        \label{quasi}
\end{figure}
In order to illustrate the importance of Coulomb corrections for quasi-elastic $(e,e')$
scattering, we shortly review the basic properties of this scattering process.
For this purpose, we envisage an electron with initial and final asymptotic four-momenta
$k^\mu_{i,f}=(\epsilon_{i,f},\vec{k}_{i,f})$, which scatters off a nucleon.
We will always set $\hbar=c=1$ in the following,
and for highly relativistic electrons we have $\epsilon_{i,f}=|\vec{k}_{i,f}|$.
Additionally, we assume that the nucleon inside the nucleus is at rest.
Neglecting interactions on the nucleon with its surrounding such that
the nucleon can be considered quasi-free, the initial and final momenta of
the nucleon are given by $p^\mu_i=(m_n,\vec{0})$ and
$p^\mu_f=(E_f,\vec{p}_f)=(m_n+\omega,\vec{k}_i-\vec{k}_f)$,
where $\omega=k^0_i-k^0_f$ is the energy
transfer and $\vec{q}=\vec{k}_i-\vec{k}_f$ the three-momentum transfer of the
electron to the nucleon. From four-momentum conservation
\begin{equation}
q^\mu=(k_i^\mu-k_f^\mu)=(p_f^\mu-p_i^\mu)
\end{equation}
we obtain from the four-momentum transfer squared $Q^2$
\begin{equation}
-Q^2=q_\mu q^\mu=2m_n^2-2m_n E_f,
\end{equation}
and consequently $\omega=(E_f-m_n)=\frac{Q^2}{2 m_n}$.
Therefore, under the simplifying assumptions made above, the $(e,e')$
scattering cross section as a function of the energy transfer for fixed electron
scattering angle $\Theta_e$ should exhibit a peak where
\begin{equation}
\omega=\frac{Q^2}{2 m_n}. \label{peakempirical}
\end{equation}

Fig. \ref{quasi} shows such a typical experimental curve from measurements
taken at Saclay \cite{Saclay}. First, one observes that the peak has
a width which is basically due to the Fermi motion of the nucleons.
Second, the peak is shifted with respect to the empirical formula
eq. (\ref{peakempirical}), which predicts $\omega_{peak} \simeq 100$ MeV, to
a value of nearly $140$ MeV. A phenomenological description of this observation
could be given within the Fermi gas model by the observation
that eq. (\ref{peakempirical}) does not take into account
that an average removal energy $\bar{E}_{rem}$ is necessary to remove a nucleon
from the nucleus which is larger than the average binding energy
$\bar{E}_{bind}$ of a nucleon inside the nucleus. E.g., for $^{208}$Pb
with $\bar{E}_{bind} \simeq 20$ MeV, a two-parameter fit for the Fermi momentum
$k_F$ and the removal energy $\bar{E}_{rem}$ leads to $k_F \simeq 265$ MeV and
$\bar{E}_{rem} \simeq 44$ MeV \cite{Whitney}. The higher value
of the removal energy also incorporates correlation effects due to the short range
interaction of the nucleons \cite{Benhar94}. However, there is a significant
non-quasielastic background present in Fig. \ref{quasi}, which if removed
would make the peak appear around $130$ MeV, a value which is not so different
from what one would expect from the binding energy, putting the
observations made above into perspective.
Furthermore, the momentum of the electron in the nuclear vicinity
is enhanced due to the attraction of the nucleus, which induces an additional
positive shift of the peak. This leads us to the idea of effective momenta.
From a classical point of view, the momentum of a highly relativistic electron
which moves virtually on a straight line is locally dependent and given by
\begin{equation}
\vec{k}_{i,f} (\vec{r}) = (k_{i,f}-V(\vec{r})) \hat{k}_{i,f},
\end{equation}
where $\hat{k}_{i,f}$ is the unit vector in direction of $\vec{k}_{i,f}$
$k_{i,f}=|\vec{k}_{i,f}|$, and $V(\vec{r})$ is the potential
energy of the electron in the electrostatic field of the nucleus.
This local change of the momentum of, e.g., the incoming particle
with momentum $\vec{k}_i=k_i \hat{k}_i$ is taken into account
by the eikonal approximation through a modification of the plane
wave part of the free wave function describing the initial state of the particle.
Defining the relativistic eikonal phase
\begin{equation}
\chi_i(\vec{r})=-\int \limits_{-\infty}^{0} V(\vec{r}+
\hat{k}_i s) ds=
-\int \limits_{-\infty}^{z} V(x,y,z') dz' \,
\end{equation}
if we choose $\vec{k}_i=k_z^i \hat{{\bf{e}}}_z$,
the free electron spinor used in PWBA calculations
\begin{equation}
\Psi_i (\vec{r})=u_{s_i}(\vec{k_i}) e^{i \vec{k}_i\vec{r}}
\end{equation}
is replaced by
\begin{equation}
\Psi_i (\vec{r})=u_{s_i}(\vec{k_i}) e^{i \vec{k}_i\vec{r}+i\chi_i(\vec{r})}
\end{equation}
in the corresponding eikonal distorted wave Born approximation (EDWBA).
$u_{s_i}(\vec{k_i})$ is the constant spinor which depends on the spin (helicity)
and momentum of the particle.
As desired, the dominant longitudinal $z$-component of the momentum $p_z$ is then
recovered via
\begin{equation}
p_z e^{i k_z^i z+i\chi_i}=-i \partial_z e^{i k_z^i z+i\chi_i}=
(k_z^i-V)e^{i k_z^i z+i\chi_i}.
\end{equation}
The final state wave function is constructed analogously by the replacement
$e^{i \vec{k}_f\vec{r}} \rightarrow e^{i \vec{k}_f\vec{r}-i\chi_f(\vec{r})}$,
where
\begin{equation}
\chi_f(\vec{r})=-\int \limits_{0}^{\infty} V(\vec{r}+
\hat{k}_f s') ds' \, .
\end{equation}
However, this approximation does not yet include the fact that also the
amplitude of the electron wave function corresponding to initial and final
asymptotic momenta $\vec{k}_{i,f}$ is modified by the attractive nucleus.
An improved version of the eikonal approximation thus should read
\begin{equation}
\Psi_{i,f}(\vec{r})=f_{i,f}^{1/2} (\vec{r})
u_{s_{i,f}}(\vec{k}_{i,f}) e^{i \vec{k}_{i,f}\vec{r}+i\chi_{i,f}(\vec{r})} ,
\end{equation}
such that the electron probability density is locally enhanced by focusing factors
$f_{i,f} (\vec{r})$.

A simpler strategy than the eikonal approximation, which will eventually lead
to the EMA and which avoids the introduction of non-planar wave functions, is to
average the locally dependent momentum over the nuclear volume, 
such that effective momenta $\vec{k}_{i,f}^{eff}$ are obtained
\begin{equation}
\vec{k}_{i,f}^{eff}=\langle \vec{k}_{i,f} (\vec{r}) \rangle=
\frac{\int \vec{k}_{i,f} (\vec{r})
\rho(\vec{r}) d^3 r}{ \int \rho(\vec{r}) d^3 r},
\end{equation}
with $\rho(\vec{r})$ representing a reasonable nuclear density profile.
If both the charge and the nuclear density are approximated by a homogeneous
distribution inside a sphere with radius $R$
\begin{equation}
\rho(\vec{r}) = \left\{ \begin{array}{ccc}
const. & : & |\vec{r} \, | \leq R  \\
0 & : &  |\vec{r} \, |>R
\end{array} \right. \quad ,
\end{equation}
then it is straightforward to show that the effective momenta are given by
\begin{equation}
\vec{k}^{eff}_{i,f} = \Bigl( k_{i,f}-\frac{4}{5} V(0) \Bigr) \hat{k}_{i,f} =
(k_{i,f}-V_{eff}) \hat{k}_{i,f}, \label{effmom}
\end{equation}
and the potential energy of an electron $V(0)$ in the center of the nucleus
is given by $V(0)=-\frac{3 \alpha Z}{2 R}$, where $\alpha=e^2/4 \pi$
is the fine structure constant and $e$ the elemental charge.
Accordingly, one can define now an effective four-momentum transfer squared
$Q^2_{eff}$. The effective potential
$V_{eff}=4 V(0)/5$ is indeed the average value of the
potential $V$ generated by the homogeneous charge distribution inside the
sphere itself. 
Eventually, since the present discussion has a phenomenological
character due to the complex and partially uncertain structure of the nuclear current,
one may modify eq. (\ref{peakempirical}) to an even more general form
\begin{equation}
\omega=\frac{Q_{eff}^2}{2 {\tilde{m}}_n} + {\tilde{E}}_{rem},
\end{equation}
where ${\tilde{m}_n}$ and ${\tilde{E}}_{rem}$ are a phenomenological
(momentum-dependent) nucleon mass and a phenomenological removal energy, respectively.
Replacing $Q^2$ by $Q^2_{eff}$ leads to an additional peak shift of
$\sim 8$ MeV in the present example.
\begin{figure}[htb]
        \centering
        \includegraphics[width=10 cm]{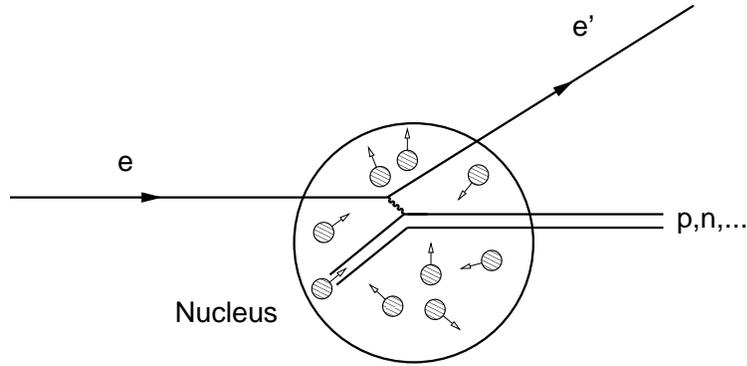}
        \caption{Quasi-elastic electron scattering off a heavy nucleus.
                 Within a strongly simplified picture, the process can
                 be viewed as scattering of the electron off the constituents
                 (mainly nucleons) of the nucleus via exchange of a `hard
                 short-range' photon.}
        \label{ee}
\end{figure}

Viewing quasi-elastic scattering as a nucleon knock-out process provides
only a poor picture of the actual physical processes taking place inside the
nucleus and for details we refer to the literature \cite{Benhar}. What is important
for the forthcoming section is the fact that the electron interacts with the nuclear
medium by exchange of photons, and that the hard scattering process can be viewed
as a {\emph{quasi-local}} process.
E.g., for $\epsilon_i=485$ MeV and $\omega=160$ MeV, the four-momentum
transfer is $Q^2=(397 \, \mbox{MeV})^2$. Taking into account that $\hbar c = 197
\, \mbox{MeV fm}$, the virtuality $Q^2$ of the exchanged photon corresponds to
a typical space-time length scale of $0.5$ fm, which is much smaller than the size of
the nucleus, as depicted in Fig. \ref{ee}. 

\section{Effective momentum approximation}
The differential cross section for single nucleon knockout is
given by \cite{Udias}
\begin{equation}
\frac{d^4 \sigma}{d \epsilon_f d\Omega_f dE_f d\Omega_f}=
\frac{4 \alpha^2}{(2 \pi)^9} \epsilon_f^2 E_f p_f
\delta(\epsilon_i \! + \! E_A \! - \! \epsilon_f \!- \! E_f
\! - \! E_{A-1}) \sum \limits_{}^{ - \! \! -} \, |W_{if}|^2 \label{crosssection},
\end{equation}
with the matrix element
\begin{equation}
W_{if}=\int \! d^3 x \!  \int d^3 y \! \int \! d^3 q \, \Bigl[
j_\mu^e(\vec{x}) \, \frac{e^{-i\vec{q} (\vec{x}-\vec{y})}}{q_\mu^2} \,
J^\mu_N(\vec{y}) \Bigr], \label{matrixelement}
\end{equation}
where $J^\mu_N(\vec{y})$ is the nucleon current obtained from
some suitable nuclear model, the $\sum \limits_{}^{ - \! \! -}$ in
eq. (\ref{crosssection}) indicates the sum (average) over final (initial)
polarizations, and $E_A$, $E_{A-1}$ is the energy of the initial and
final nucleus, respectively.
In the PWBA, the electron current is given by
\begin{equation}
j^\mu(\vec{x})=
\bar{u}_{s_f}(\vec{k}_f) \gamma^\mu u_{s_i}(\vec{k}_i)
e^{i \vec{k}_i \vec{r}-i \vec{k}_f \vec{r}},
\end{equation}
where ${u}_{s_i},{u}_{s_f}$ are initial/final state
plane wave electron spinors corresponding to the initial/final electron
momentum $\vec{k}_{i,f}$ and spin $s_{i,f}$.
In the DWBA, exact solutions of the Dirac equation are used for electrons instead
of plane waves. The usual procedure to calculate the inclusive $(e,e')$ cross
section is to sum over all the individual nucleon knockout cross sections for
all protons and neutrons in the nucleus under consideration.

The basic idea of the effective momentum approximation (EMA)
is to describe the electron wave functions by modified plane waves
\begin{displaymath}
e^{i \vec{k}_{i,f} \vec{r}} \rightarrow \frac{{k}_{i,f}'}
{k_{i,f}} e^{i \vec{k}_{i,f}'' \vec{r}} \, , \label{EMA}
\end{displaymath}
which account for the enhanced electron density and momentum
in the nuclear region. Here, $k_{i,f}'$ and $k_{i,f}''$ denote effective
momenta which need not necessarily be identical. It will be one of the main
results of the forthcoming section, that for high electron energies and
the electrostatic potential of a homogeneously charged sphere,
$k_{i,f}'=k_{i,f}''=k_{i,f}^{eff}$ is indeed fulfilled, i.e., at high energies, the
effective (average) focusing factor is given by
\begin{equation}
f_{i,f}=\frac{\int f_{i,f} (\vec{r})
\rho(\vec{r}) d^3 r}{ \int \rho(\vec{r}) d^3 r}=\Biggl(\frac{k_{i,f}^{eff}}{k_{i,f}}
\Biggr)^2.
\end{equation}
We will therefore identify the $k_{i,f}'=k_{i,f}''=k_{i,f}^{eff}$ in the sequel.
When the exact wave functions appearing in the matrix element eq. (\ref{matrixelement})
are replaced by the corresponding effective wave functions, the momentum integral
in eq. (\ref{matrixelement}) can be trivially performed and replaced basically
by a constant factor $1/q_{\mu,eff}^2=-1/Q_{eff}^2$.
This expresses the fact that the virtual photon emitted by the electron
is actually harder than if no attractive potential were present,
since the electron is accelerated to higher momenta in
the nuclear vicinity, and $1/q_{\mu,eff}^2$ is the photon virtuality averaged over
the nuclear volume.

Note that the reason why the replacement of the locally
dependent wave function amplitudes and momenta by effective values makes
sense is rooted in the local character of the scattering process mentioned above.
E.g., if the virtual photon would propagate over distances comparable to
the size of the nucleus, then nucleons could also be knocked by photons
which were emitted outside the nucleus, such that an averaging of
the focusing and the local momenta inside the nuclear interior would not make sense.
The correct mathematical counterpart of this pictorial description can be found
in \cite{Knoll}. For the EMA to hold, it is mandatory that the
wave lengths of the electron and the virtual photon are significantly smaller
than the nuclear radius, i.e. $\epsilon_f > 200$ MeV and $Q^2>(200 \, \mbox{MeV})^2$
corresponding to a length scale of $1$ fm should be required for $^{208}$Pb
\cite{Jin,Aste3}.
Note also that the enhancement of the wave function amplitudes
is not very large at high energies. One can write
\begin{equation}
f^{1/2}_{i,f}(\vec{r})=1+\delta_{i,f}(\vec{r}), \quad |\delta(\vec{r})| \ll 1,
\end{equation}
and we may therefore neglect higher order terms in the $\delta$'s in formal
expressions, like
\begin{equation}
\langle f_i^{1/2}(\vec{r}) f_f^{1/2}(\vec{r}) \rangle =
\langle (1+\delta_{i,f}(\vec{r})) (1+\delta_{i,f}(\vec{r})) \rangle \simeq
\langle f_i^{1/2} (\vec{r})\rangle \langle f_f^{1/2}(\vec{r}) \rangle,
\end{equation}
and one may equate expressions like
\begin{equation}
f_{i,f} (\vec{r}) = (1 + \delta_{i,f}(\vec{r}))^2 \simeq 1+ 2 \delta_{i,f}(\vec{r}).
\end{equation}

It is instructive to calculate the impact of the focusing factors on
the size of the cross section for a typical example. 
E.g., if we consider electron scattering off $^{208}$Pb for $|k_{i}|=485$ MeV/c
and $\omega=100$ MeV, we have $V(0)=-25$ MeV and $4V(0)/5=-20$ MeV such that
$k_{i}^{eff}$ is given by $(485+20)$ MeV, and
$k_{f}^{eff}=(385+20)$ MeV.
The focusing factors enter the cross section both linearly via the matrix
element squared, enhancing the cross section by a factor of
$(k_{i}^{eff}/k_i)^2(k_{f}^{eff}/k_f)^2=1.2$.

There are two different, but equivalent strategies to calculate cross
sections in the EMA framework.
First, the EMA cross section can be calculated
by replacing the electron momenta $\vec{k}_{i,f}$ by the effective momenta
$\vec{k}_{i,f}^{eff}$ in the (theoretical) expression for the
quasi-elastic scattering cross section (accordingly, the energies $\epsilon_{i,f}$
must be replaced by $|\vec{k}^{eff}_{i,f}|$).
The cross section obtained this way must be multiplied
subsequently by the factor $(k_{i}'/k_i)^2$
which accounts for the focusing of the incoming electron wave
in the nuclear center. 
The focusing factor $(k_{f}'/k_f)^2$ for the scattered
electron is already contained in the artificially enhanced phase space
factor of the final state electron, if $k'_f=k''_f$ is presumed.
Second, the cross section for
inclusive quasi-elastic electron scattering can also be written by
the help of the total response function $S_{tot}$ as
\begin{equation}
\frac{d^2 \sigma_{_{PWBA}}}{ d \Omega_f d\epsilon_{f}}=
\sigma_{Mott} \times S_{tot}(|\vec{q} \, |,\omega,\Theta_e),
\label{Mott1}
\end{equation}
where the Mott cross section is given by ($q_\mu^4=Q^4$)
\begin{equation}
\sigma_{Mott}=4 \alpha^2 \cos^2(\Theta_e/2) \epsilon_f^2/q_\mu^4.
\label{Mott2}
\end{equation}
The Mott cross section remains unchanged
when it gets multiplied by the EMA focusing factors
and the momentum transfer $q_\mu^4$ is replaced by its
corresponding effective value. A short calculation shows indeed that
($\epsilon_{i,f} \gg m$)
\begin{equation}
\frac{Q_{eff}^2}{Q^2}=\frac{k_i^{eff} k_f^{eff}}{k_i k_f}, \quad
\frac{f_i (\epsilon_{f}^{eff})^2}{Q_{eff}^4}=\frac{\epsilon_f^2}{Q^4}.
\end{equation}
Therefore, the EMA cross section can also be obtained
from (\ref{Mott1}) by leaving the Mott cross section
unchanged and by replacing $S_{tot}(|\vec{q} \, |,\omega,\Theta_e)$
by the effective value
\begin{equation}
S_{tot}(|\vec{q}_{eff}|, \omega,\Theta_e)=
S_{tot}(|\vec{k}_i^{eff}-\vec{k}^{eff}_f|, \omega,\Theta_e),
\end{equation}
since the effect of replacing $q_\mu^4$ by its effective value in
the Mott cross section is to exactly divide away the initial
state focusing factor and the final state focusing factor which
is generated by the replacement
$\epsilon_f \rightarrow \epsilon_f^{eff}=|k_f^{eff}|$.

\section{The classical focusing factor} 
\begin{figure}
\begin{center}
        \includegraphics[width=10.8cm]{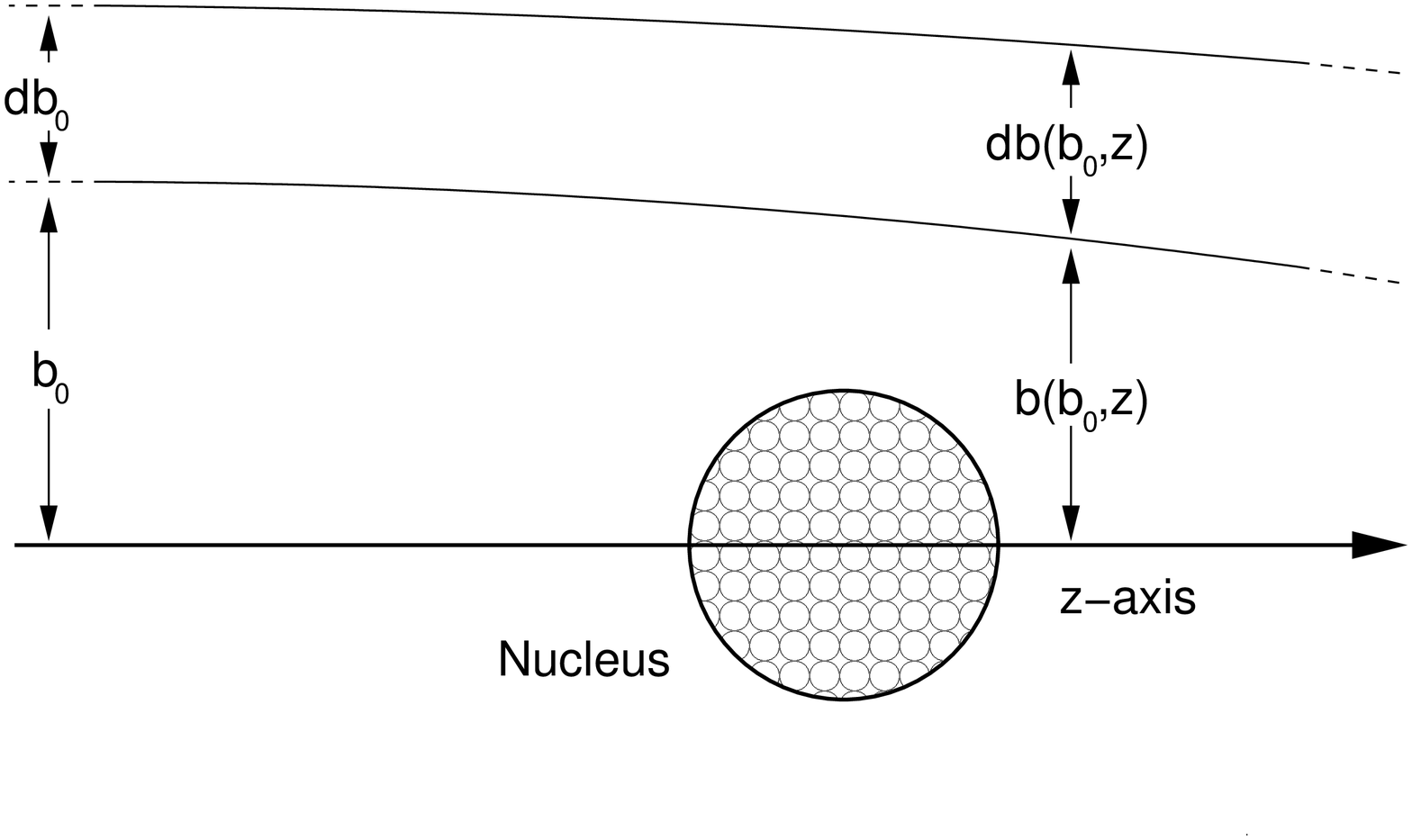}
        \caption{Electrons incident on an
         nucleus with impact parameter $b_0$.}
        \label{figfoc}
\end{center}
\end{figure}
\begin{figure}
\begin{center}
        \includegraphics[width=7.8cm]{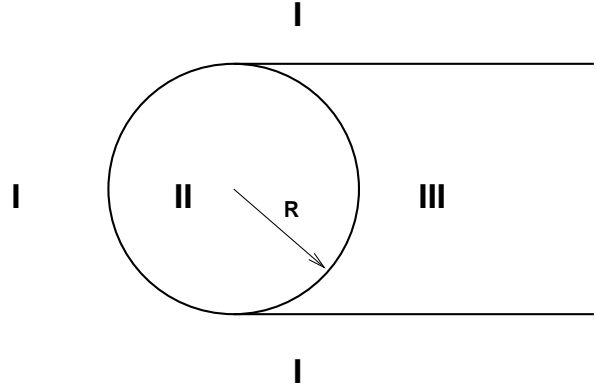}
        \caption{The three different regions according to the case distinction
                 in the text.}
        \label{regions}
\end{center}
\end{figure}

The focusing factor can be derived approximately from a classical
toy model according to Figs. \ref{figfoc} and \ref{regions}.
We consider the trajectories of an ensemble of highly relativistic
particles approaching a nucleus located in the center of the $(b,z)$-coordinate system.
The particles shall move in $z$-direction with equal velocity and with an
asymptotic impact parameter in the range between $b_0$ and $b_0+db_0$.
The longitudinal velocity of the particles can be taken
as the speed of light, since the particles are
highly relativistic and changes of the velocity in transverse
direction and the kinetic energy cause a negligible second-order effect to the
longitudinal component.
Therefore, one may adopt the straight-line approximation by
setting $z(t)=ct=t$, $r(t)=\sqrt{b_0^2+t^2}$, and
the impact parameter will be considered constant at certain stages
of our calculation. Due to the attractive nucleus, the original
impact parameter $b_0$ is reduced to $b(b_0,z)$ as a particle moves
along its trajectory, such that the particle density at $z$ is
increased by a focusing factor $f$ which is given by the ratio of the area
of two annuli with radii
$b_0, b_0+db_0$ and $b(b_0,z), b(b_0+db_0,z)$:
\begin{equation}
f^{-1}(b_0,z) \simeq f^{-1}(b(b_0,z),z)=\frac{\partial b(b_0,z)}{\partial b_0}
\frac{b(b_0,z)}{b_0}. \label{focusing_general}
\end{equation}

In the following, we will calculate $b(b_0,z)$ for an electron in the potential
of a homogeneously charged sphere with radius $R$ and charge
$eZ$, given by
\begin{equation}
V_{hom}(r) = \left\{ \begin{array}{ccc}
-\frac{\alpha Z}{R} \Bigl( \frac{3}{2}- \frac{r^2}{2 R^2} \Bigr) & : & r \leq R  \\
-\frac{\alpha Z}{r} & : &  r>R
\end{array} \right. \quad .
\end{equation}
since the potential of a homogeneously charged sphere provides a simple but quite
realistic model for the electromagnetic field of a heavy nucleus like, e.g.,
$^{208}$Pb, where one has $R \simeq 7.1 \, \mbox{fm}$ and $Z=82$.

The force $f_t$ acting on the particle in transverse direction is
given by
\begin{equation}
f_t = -\frac{b}{r} \frac{\partial V_{hom}(r)}{\partial r} \sim
-\frac{b_0}{r} \frac{\partial V_{hom}(r)}{\partial r}, \label{transverse_force}
\end{equation}
In eqns. (\ref{focusing_general}) and (\ref{transverse_force}),
we made use of the straight-line assumption by
replacing $b$ by $b_0$.
Also in the forthcoming, we will sometimes replace $b$ by $b_0$ or use these
two quantities synonymously where such a substitution is adequate.
To calculate the transverse acceleration of the particle due to the
attractive Coulomb field, we distinct three cases (see Fig. \ref{regions}).
In the first case, we consider the region where the particle
moves solely in the $1/r$-field according to the straight-line approximation,
i.e. where we have $b_0>R$ (or $b_0<-R$, if we formally allow negative impact
parameters), plus the region with $b_0<R$ and $z<-\sqrt{R^2-b_0^2}$.
The transverse force is then given by
\begin{equation}
f_t^I=-\frac{\alpha Z b_0}{r^3}.
\end{equation}
Correspondingly, we obtain for the transverse acceleration, taking into
account that the "transverse mass" of a relativistic particle
is given by its energy $E$, which is also considered as constant:
\begin{equation}
a_t^I=\dot{v}_t^I=
-\frac{\alpha Z}{E}\frac{b_0}{r^3}=
-\frac{\alpha Z}{E}\frac{b_0}{\sqrt{t^2+b_0^2}^3},
\end{equation}
where $r$ is the distance of the particle from the nuclear center, and
\begin{equation}
v_t^I(b_0,t)=\int \limits_{-\infty}^{t} a_t^I (b_0,t') dt'=
-\frac{\alpha Z}{E} \frac{t+\sqrt{t^2+b_0^2}}
{b_0 \sqrt{t^2+b_0^2}}=
-\frac{\alpha Z}{E} \frac{t+r}
{b_0 r}. \label{velocity}
\end{equation}
Note that from $v_t^I(t \rightarrow \infty)=-\frac{2\alpha Z}{E}$
we obtain for a pure $1/r$-Coulomb field
the well-known transverse momentum transfer
\begin{equation}
\Delta k_t = \frac{2 \alpha Z}{b_0}.
\end{equation}
Furthermore, we obtain from (\ref{velocity})
\begin{equation}
b^I(b_0,z)=b_0+\int \limits_{-\infty}^{z} v_t^I(b_0,t') dt'
=b_0-\frac{\alpha Z}{E} \frac{z+r}{b_0}. \label{classbI}
\end{equation}
A short calculation yields the focusing factor
\begin{equation}
f^I=\Biggl[ 1-\frac{\alpha Z}{E} \Biggl( \frac{1}{r}+\frac{z+r}{b_0^2}
\Biggr) \Biggr]^{-1} \Biggl[ 1-\frac{\alpha Z}{E} \Biggl( \frac{z+r}{b_0^2}
\Biggr) \Biggr]^{-1}.
\label{focus2}
\end{equation}
Since we are interested in the high energy behavior, we keep in eq. (\ref{focus2})
only the relevant zeroth- and first-order terms in $\alpha Z/E$.
For the focusing factor in region I we obtain the simple result
\begin{equation}
f^I(b_0,z)=1+\frac{\alpha Z}{E} \frac{1}{r}. \label{classicalfocus1}
\end{equation}
The calculation for regions II and III are a bit more involved, but
can be performed along the same lines as above.

We calculate now the focusing inside the charged sphere (region II),
which gets traversed by particles with $b_0<R$.
Inside the sphere, the transverse acceleration of the particles is due
to the harmonic oscillator potential generated by the homogeneous charge
distribution. Correspondingly, we have
\begin{equation}
a_t^{II} = -\frac{b}{r} \frac{\alpha Z}{E} \frac{r}{R^3}
\sim -\frac{\alpha Z}{E} \frac{b_0}{R^3},
\end{equation}
and for the transverse distance from the $z$-axis we obtain after a short
calculation
\begin{equation}
b^{II} (b_0,z)=b_0-\frac{\alpha Z}{E} \Biggl[
\frac{R-\tilde{R}}{b_0} + \frac{\frac{b_0}{2} \Bigl(
z+\tilde{R} \Bigr)^2}{R^3} +
\frac{\Bigl( R - \tilde{R} \Bigr) \Bigl( \tilde{R}+z \Bigr)}
{b_0 R} \Biggr].
\end{equation}
Above, we have introduced the abbreviation $\tilde{R}=\sqrt{R^2-b_0^2}$.
Note that the first term in the bracket above describes the transverse
shift of the particles when they arrive on the surface of the charged sphere according to
eq. (\ref{classbI}), where $z=-\tilde{R}$. The second term is due to the
transverse acceleration of the particles inside the sphere, and
the last term in the expression above is generated by the transverse
velocity $v_t^I (b_0,-\tilde{R})$
which is reached by the particles when they cross the border of the sphere.
A straightforward calculation leads to the following result for the focusing
factor at first order in $\alpha Z/E$:
\begin{equation}
f^{II}(b_0,z)=1 +\frac{\alpha Z}{ER} \Bigl(3+ 3 \frac{z \tilde{R}}{R^2} +
\frac{z^2}{R^2}-2b_0^2/R^2 \Bigr).
\end{equation}
In the center of the nucleus, the particle density is enhanced by
a factor
\begin{equation}
f^{II}(0)=f^{II}(b_0=0,z=0)=1+\frac{3 \alpha Z}{ER},
\end{equation}
however, the average focusing factor inside the sphere is given by
the volume integral
\begin{equation}
\bar{f}^{II}= \frac{\int \limits_{b_0^2+z^2<R^2} f^{II}(b_0,z) dV}{\frac{4 \pi}{3} R^3}
=\frac{4}{5} f^{II}(0),
\end{equation}
i.e. one obtains the focusing factor used in the effective
momentum approximation, where the increased particle probability density near the
nucleus is taken into account by multiplying the particle's Dirac wave
function by a suitable factor $f^{1/2} \simeq (\bar{f}^{II})^{1/2}$.

Finally, we consider the `shadow region' of the nucleus (region III).
We calculate first the transverse velocity of the particle when it
arrives in region III in three steps.
First, the particle moves inside the $1/r$-field and reaches a transverse
velocity
\begin{equation}
v_1=v_t^I(b_0,-\tilde{R})=-\frac{\alpha Z}{E} \frac{R-\tilde{R}}{b_0 R}
\end{equation}
at the surface where it enters the sphere.
Inside the sphere, the particle undergoes a constant transverse acceleration
$a_t^{II}$ for $t \! \in \! [-\tilde{R},\tilde{R}]$. Therefore,
the particle gains an additional velocity
\begin{equation}
v_2=-\frac{\alpha Z}{E} \frac{2 b_0 \tilde{R}}{R^3}.
\end{equation} 
In the downstream region III, the particle is again moving in the field of a
point-like charge, and the transverse acceleration is given by
\begin{equation}
a_t^{III}=-\frac{\alpha Z}{E} \frac{b_0}{\sqrt{t^2+b_0^2}^3},
\end{equation}
such that we end up with ($r=\sqrt(t^2+b_0^2)$)
\begin{displaymath}
v_t^{III}=v_1+v_2+\int \limits_{\tilde{R}}^{t} a_t^{III}(b_0,t') dt'=
\end{displaymath}
\begin{displaymath}
-\frac{\alpha Z}{E} \Biggl[
-\frac{\tilde{R}}{b_0 R} +\frac{1}{b_0} +\frac{2 b_0 \tilde{R}}{R^3}
\Biggr]-\frac{\alpha Z}{E} \Biggl[ \frac{t+r}{b_0 r}
-\frac{\tilde{R}}{b_0 R} - \frac{1}{b_0} \Biggr]=
\end{displaymath}
\begin{equation}
-\frac{\alpha Z}{E} \Biggl[ \frac{t+r}{b_0 r}
-\frac{2 \tilde{R}}{b_0 R} + \frac{2 b_0 \tilde{R}}{R^3} \Biggr].
\end{equation}
The transverse distance of the particles from the $z$-axis in region III
is therefore given by
\begin{displaymath}
b^{III}(b_0,z)=b^{II}(b_0,\tilde{R})+\int \limits_{\tilde{R}}^{z}
v_t^{III}(b_0,t) dt=
\end{displaymath}
\begin{displaymath}
b_0-\frac{\alpha Z}{E} \Biggl[ -\frac{R}{b_0} + \frac{4 b_0}{R} +
\frac{\tilde{R}}{b_0}-\frac{2 b_0^3}{R^3} \Biggr]+
\int \limits_{\tilde{R}}^{z}
v_t^{III}(b_0,t) dt=
\end{displaymath}
\begin{equation}
b_0-\frac{\alpha Z}{E} \Biggl[ \frac{z}{b_0}+\frac{2 b_0 z \tilde{R}}{R^3}
+\frac{\sqrt{z^2+b_0^2}}{b_0}-\frac{2 z \tilde{R}}{b_0 R} \Biggr].
\label{bregionIII}
\end{equation}
For the focusing factor we obtain from eq. (\ref{bregionIII})
\begin{equation}
f^{III}(b_0,z)=1+\frac{\alpha Z}{E} \Biggl[ \frac{6z}{R \tilde{R}}
+\frac{1}{\sqrt{z^2+b_0^2}}-\frac{6z b_0^2}{R^3 \tilde{R}} \Biggr].
\end{equation}

We finally summarize the results as follows.
An attractive nucleus modeled by a homogeneously charged sphere
acts like a focusing lens on an ensemble of classical particles incident
on the nucleus with impact parameter $b_0$ on quasi-straight trajectories
parallel to the $z$-axis. For highly relativistic particles,
the particle density is enhanced by a focusing factor
\begin{equation}
f(b_0,z)=1+ \frac{\alpha Z}{E} \Phi(b_0,z)
\end{equation}
with ($r=\sqrt(b_0^2+z^2)$)
\begin{equation}
\Phi(b_0,z) = \left\{ \begin{array}{ccc}
\frac{1}{r} & : & I \\
\frac{3}{R}+ 3 \frac{z \tilde{R}}{R^3} +
\frac{z^2}{R^3}-2b_0^2/R^3  & : &  II \\
\frac{1}{r}+\frac{6z}{R \tilde{R}}
-\frac{6z b_0^2}{R^3 \tilde{R}} & : & III
\end{array} \right. \quad . \label{fullfoc}
\end{equation}
The typical deviation of the focusing from unity in the nuclear interior
is of the order of $3 \alpha Z/ER$ or $V_{hom}(0)/E$; this ratio should be considered
as the expansion parameter for higher order corrections to the focusing, which
become irrelevant at high energies.

\begin{figure}
\begin{center}
  \centerline{
    \mbox{\includegraphics[width=3.00in]{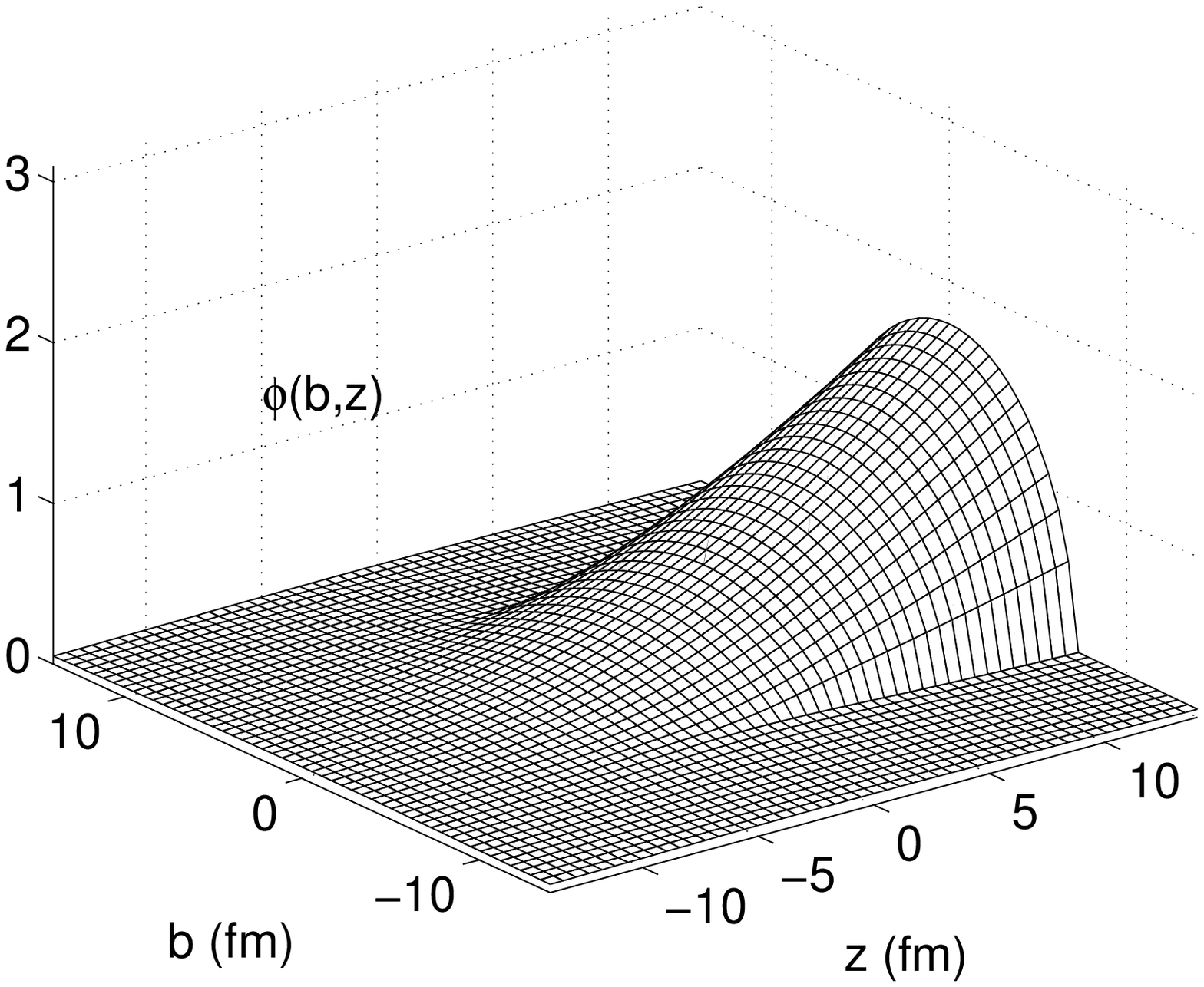}}
    \mbox{\includegraphics[width=3.00in]{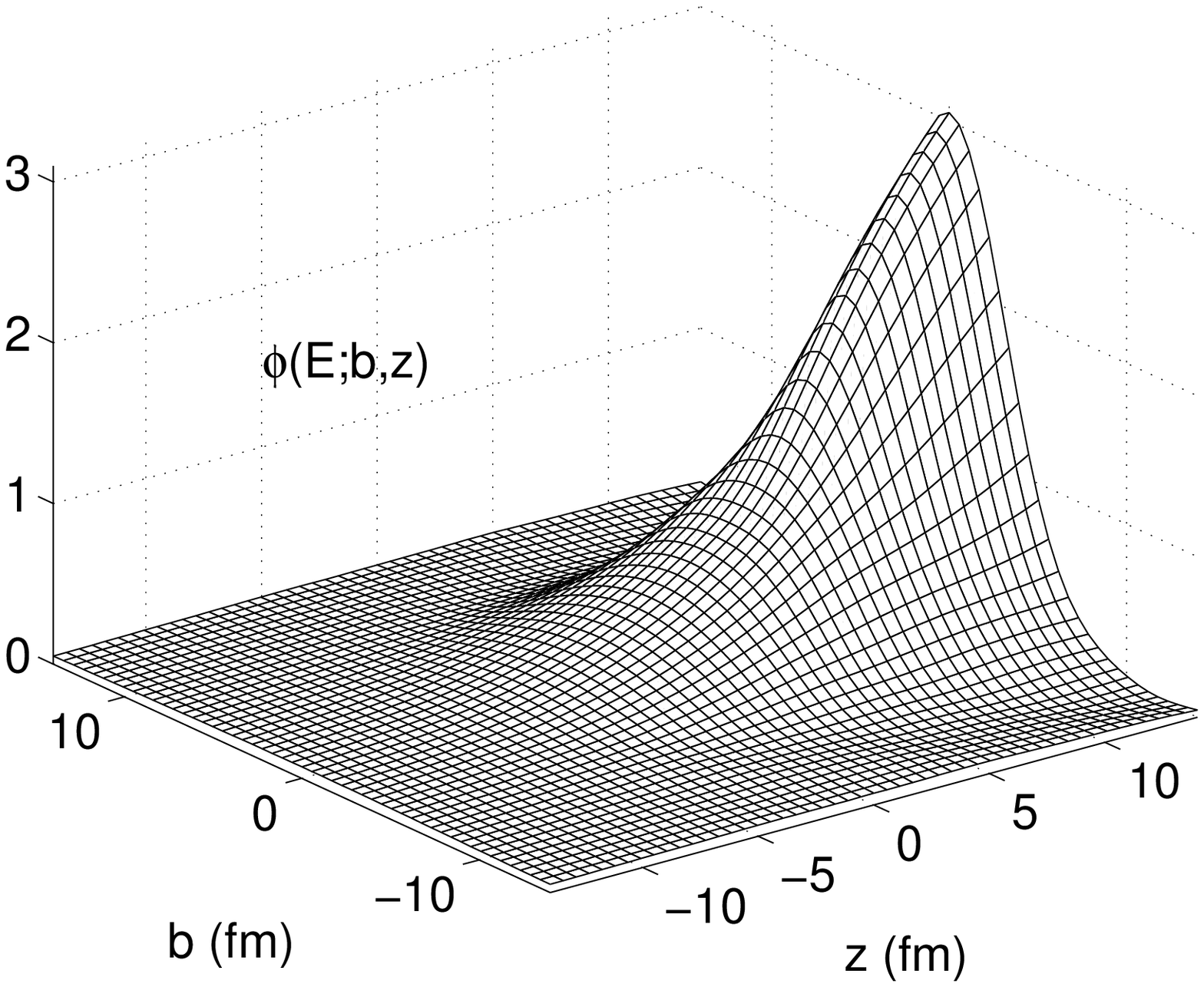}}
  }
        \caption{Left: The function $\Phi$ describing the focusing effect
                 of the attractive potential of a homogeneously charged
                 sphere on an ensemble of highly relativistic particles.
                 In order to symmetrize the figure, the impact parameter
                 $b$ can also be negative. Parameter values typical for a
                 $^{208} \, \mbox{Pb}$ nucleus have been used
                 ($R=7.1 \, \mbox{fm}$, $Z=82$). Right: Focusing function
                 $\Phi(E;b,z)$ for an electron incident on the potential
                 with $E=200 \, \mbox{MeV}$ and positive helicity.
                 The left and the right plot agree well inside the sphere, however,
                 the wave focusing is clearly larger on the rear side of the nucleus
                 than in the classical highly relativistic case.}
        \label{classical}
\end{center}
\end{figure}

\begin{figure}
\begin{center}
        \includegraphics[width=8.5cm]{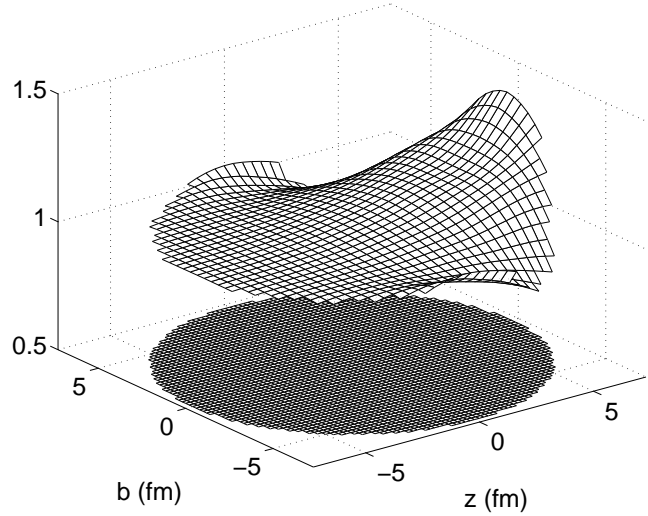}
        \caption{The ratio $\Phi(E,b,z)/
        \Phi(b,z)$ for $E=200 \, \mbox{MeV}$ in region II.}
        \label{ratio}
\end{center}
\end{figure}

\begin{figure}
\begin{center}
  \centerline{
    \mbox{\includegraphics[width=3.00in]{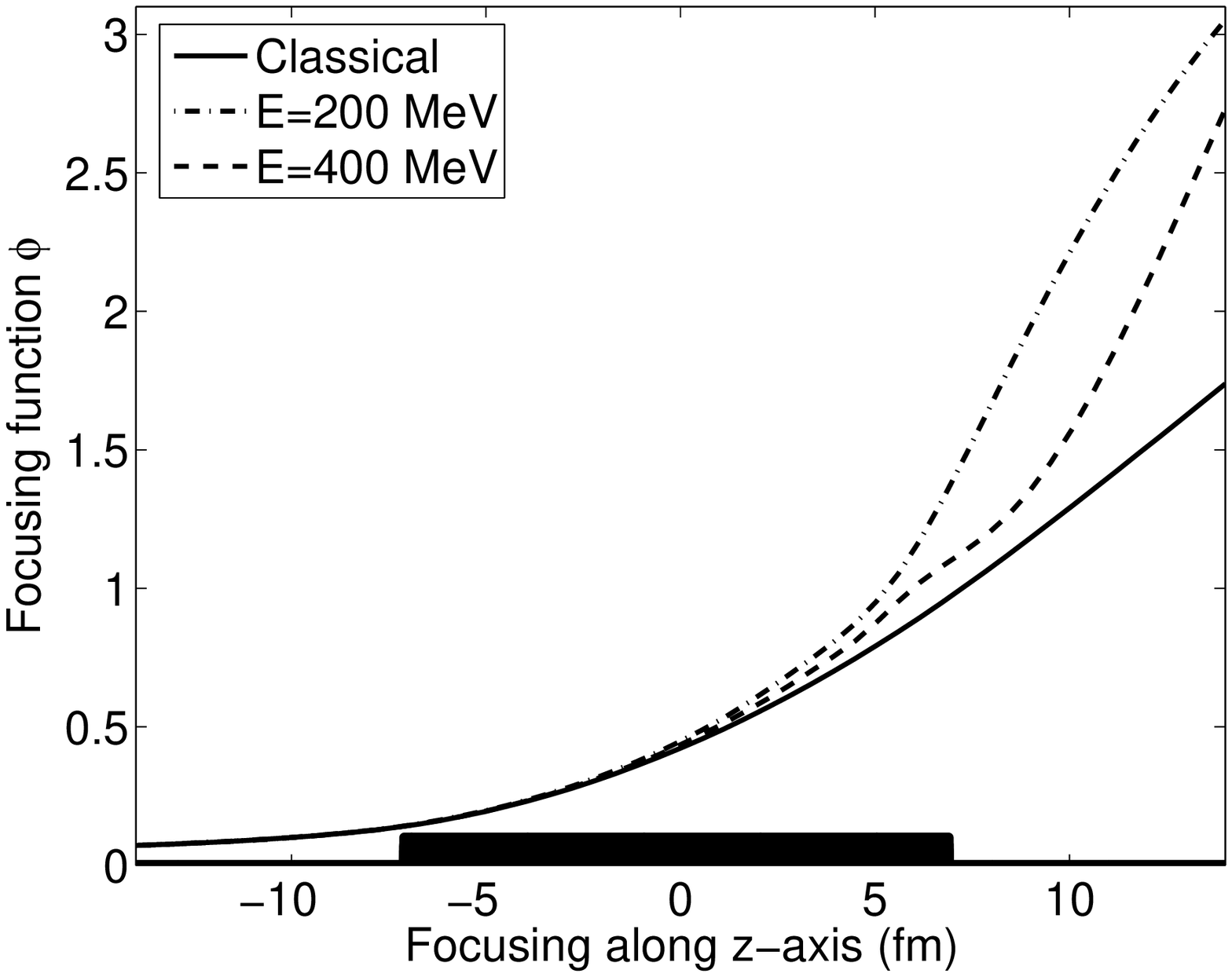}}
    \mbox{\includegraphics[width=3.00in]{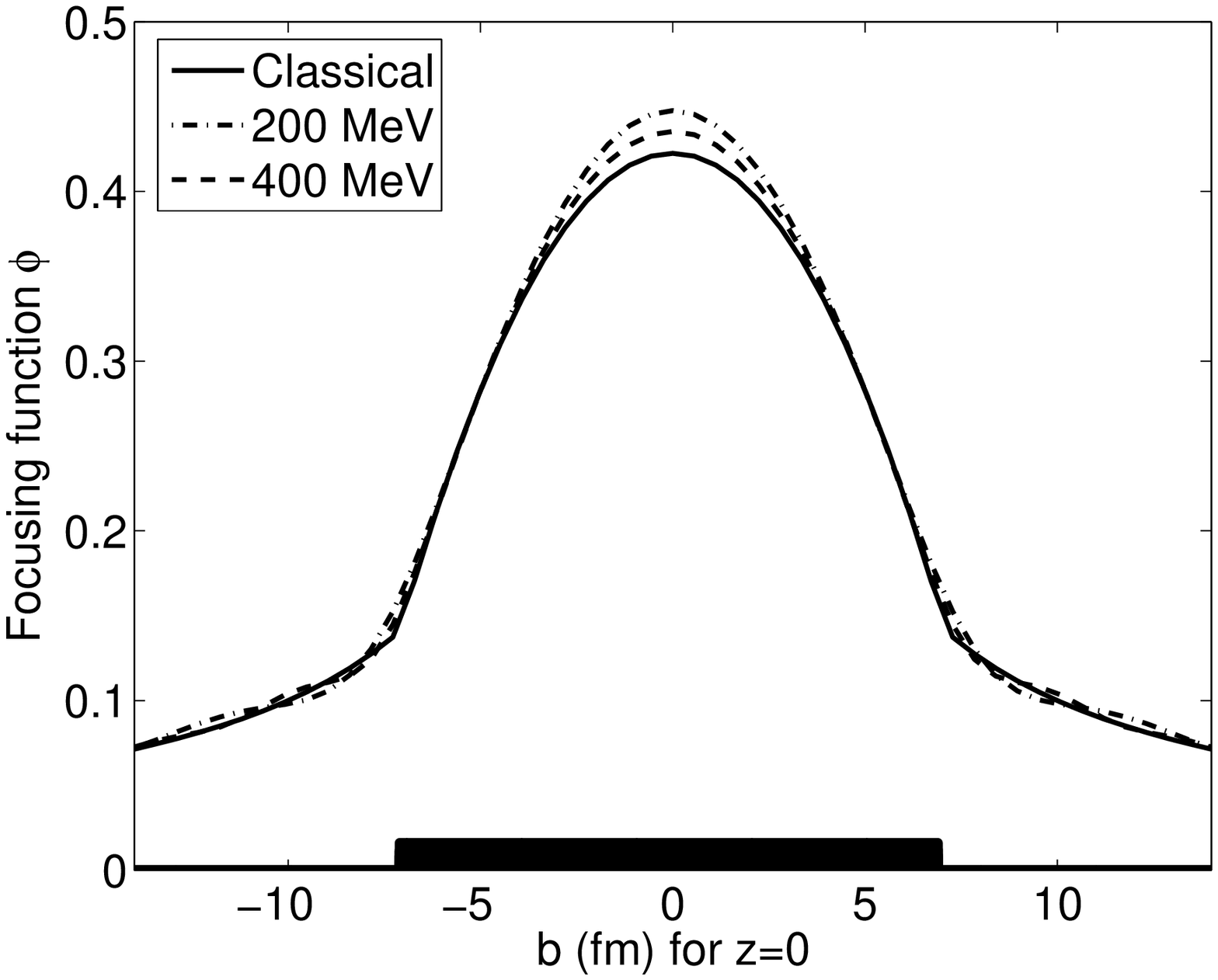}}
  }
        \caption{Left: Longitudinal focusing of the classical particle density 
                 $\Phi(b=0,z)$ versus the focusing of the Dirac wave function
                 $\Phi(E,b=0,z)$ for two different electron energies $E=200$ MeV
                 and $E=400$ MeV.
                 Right: Corresponding transverse focusing $\Phi(b,z=0)$. The bar at the
                 bottom depicts the extension of the nucleus.}
        \label{longitrans}
\end{center}
\end{figure}

Fig. \ref{classical} shows a surface plot of the universal function
$\Phi(b_0,z)$.
In order to compare the classical focusing to the results obtained by
solving the Dirac equation exactly \cite{Aste3,TBR83,Pauli}, we define
\begin{equation}
\Phi(E;b,z):=\frac{E}{\alpha Z} (\rho(E;b,z)-1),
\end{equation}
where $\rho(E;b,z)=\Psi(E;\vec{r})^\dagger \Psi(E;\vec{r})$ is the axially
symmetric probability density of the Dirac wave function
$\Psi(E;\vec{r})$ of an electron incident with asymptotic momentum $\vec{k}=
\sqrt{E^2-m^2} \hat{z} \simeq E \hat{z}$ and spin parallel to the $z$-axis.
Results are shown in Figs. \ref{classical} and \ref{longitrans}.
The Dirac density indeed approaches the classical
limit for high electron energies. Note that the focusing of the Dirac
wave function is clearly underestimated by the classical high-energy
approximation in region III, where the straight-line assumption starts to
break down.
However, in the case of quasi-elastic electron
scattering, the relevant region is the interior of the nucleus, where
the focusing is described in a satisfactory way for electron energies
above $200$ MeV as shown in Figs. \ref{classical}, \ref{ratio} and
\ref{longitrans}.

We note that Knoll \cite{Knoll} derived the focusing effect from a high energy
partial wave expansion, following previous results given
by Lenz and Rosenfelder \cite{Lenz,Rosenfelder}.
For the incoming particle wave expanded around the center of
the nucleus he obtained
\begin{displaymath}
\Psi_i(\vec{r})=e^{i \delta_i} (\tilde{k}_i/k_i) e^{i \vec{\tilde{k}}_i \vec{r}} \times
\end{displaymath}
\begin{equation}
\{1+a_1r^2-2a_2\vec{\tilde{k}}_i\vec{r}+i a_1 r^2 \vec{\tilde{k}}_i \vec{r}+ia_2 [
(\vec{\tilde{k}}_i \times \vec{r})^2+
\vec{\sigma} (\vec{\tilde{k}}_i \times \vec{r})] \} u_{s_i}(\vec{k}_i),
\label{expknoll}
\end{equation}
where $\delta_i$ is a phase, $\vec{\tilde{k}}_i$ is an effective momentum
parallel to $\vec{k}_i$ calculated by using the central potential value
$\tilde{k}_i=k_i-V_{hom}(0)$, and $\vec{\sigma}$ acts on the spinor
$u_{s_i}(\vec{k}_i)$ to describe spin dependent effects, which are negligible
in our cases of interest with definite helicity.
An analogous equation holds for the distortion of the outgoing wave.
The parameters $a_{1,2}$ depend on the shape of the
potential. For a homogeneously charged sphere with radius $R$
they are given by
\begin{equation}
a_1=-\frac{\alpha Z}{6 \tilde{k_i} R^3} , \quad 
a_2=-\frac{3 \alpha Z}{4 {\tilde{k}_i}^2 R^2} \label{para} ,
\end{equation}
and the central potential value is given by $V_{hom} (0)=-\frac{3}{2} \frac{\alpha Z}{R}$.
The increase of the amplitude of the wave while passing
through the nucleus is described
mainly by the $-2a_2\vec{\tilde{k}}_i \vec{r}$-term, the
$a_1 r^2$-term accounts for a decrease of the focusing
also in transverse direction.
Performing the replacements $r^2=b^2+z^2$, $\vec{\tilde{k}}_i \vec{r}= \tilde{k}_i z$
and $|\vec{\tilde{k}}_i \times \vec{r}|= \tilde{k}_i b$, one obtains from eq.
(\ref{expknoll})
for a particle with spin parallel to the momentum and energy
$E=k_i$ up to second order in $b$ and $z$ and first order in $\alpha Z$
\begin{equation}
\Psi_i(\vec{r})^\dagger  \Psi_i(\vec{r}) \simeq \Biggl[
1+\frac{\alpha Z}{ER} \Biggl( 3 + \frac{3z}{R}+\frac{z^2+b^2}{3 R^2} \Biggr) \Biggr]
u_{s_i}(\vec{k}_i)^\dagger u_{s_i}(\vec{k}_i),
\end{equation}
i.e. the correct linear term in $z$ in eq. (\ref{fullfoc}) is recovered, however,
the transverse decay of the focusing is strongly suppressed in the $V_{eff}=4 V(0)/5$
expansion given by Knoll, which is therefore not suited to describe
the focusing for $b \neq 0$. The use of the expansion eq. (\ref{expknoll})
is the reason why the $(e,e')$ cross sections in \cite{Aste1} are overestimated.

Calculations using exact Dirac wave function show that the effective 
electron momenta are very well described by an effective potential
$V_{eff}=4 V(0)/5$ \cite{Aste3}. Furthermore, our findings indicate that
also the focusing can be described in an accurate way by the same effective
potential value. This demonstrates the validity of the EMA as a valuable tool
for the description of Coulomb distortion effects. One may observe
from Fig. \ref{longitrans} that the exact focusing for finite energies
is slightly larger than in the high energy limit.
This leads to a minor amplification of the DWBA cross
section compared to the EMA result. Exact calculations, which will be presented
in a forthcoming paper, show that this effect is only of the order of
$2$\% in the region of the quasielastic peak for typical kinematical values
used in experiments, e.g. for an initial electron energy of
$\epsilon_i=485$ MeV and scattering angle
$\Theta_e=60^o$, or $\epsilon_i=310$ MeV and $\Theta_e=143^o$.

\section{Final remarks and conclusions}
The high energy trajectory of a charged classical particle moving
in the field of a homogeneously charged field was investigated and related
through the quantum-classical correspondence principle to
the probability density of exact continuum wave functions obtained as
solutions of the Dirac equation. As a result, a universal function $\Phi$ was
found which allows to describe the high energy behavior of
the amplitude of Dirac electron wave functions with definite helicity.
The focusing in the downstream side of the charge distribution converges slowly
towards the high-energy limit described by $\Phi$, however, the universal
function $\Phi$ provides an accurate description of the focusing inside
the charged space region, which can be considered as a model for the
charge distribution of a heavy nucleus. As a consequence, it is found
that both the effective (average) momenta and the average focusing can
be described by a common effective potential $V_{eff}=4 V(0)/5$ in the
case of a homogeneously charged sphere, despite the fact that the local
classical momenta exhibit the same spherical symmetry as the
electrostatic potential, whereas the axially symmetric focusing is smaller
in the upstream side and larger in the downstream side of the nucleus.

Our findings establish the role of the EMA as a valuable semiclassical
method for the analysis of Coulomb corrections in $(e,e')$ scattering.
They also indicate that the analysis of experimental data based on calculations
of Kim {\emph{et al.}} should be revisited \cite{Kosik,JourdanCoulomb}, and support the
strategy in previous works concerning the extraction of the longitudinal and
transverse response functions in medium-weight and heavy nuclei
\cite{MezianiCoulomb}. However, it is also advisable to await new experimental
data which will hopefully be accessible in the near future \cite{jlab}.

We finally remark that Baker investigated also the second-order eikonal approximation
for potential scattering in the non-relativistic case \cite{Baker},
finding thereby an expression for the focusing factor of continuum Schr\"odinger wave
functions. For the focusing in the center of a spherically symmetric potential,
one finds (see eq. (23) in \cite{Baker})
\begin{equation}
f^{1/2}(0)=f^{1/2}(b=0,z=0) \simeq 1-\frac{V(0)}{2 k v}, \quad \mbox{or} \quad
f(0) \simeq 1-\frac{V(0)}{k v}
\end{equation}
where $k$ is the asymptotic momentum and $v$ the velocity of the particle.
Roughly speaking, the approximation is valid if the kinetic energy of the
particle is larger than the depth of the disturbing potential
$m \gg E_{kin}=E-m \gg V(0)$, and the wave length of the particle $\sim 2 \pi/k$
should be significantly smaller than the extension of the potential.
For the classical particle momentum in the center of the potential $k(0)$ one has
non-relativistically
\begin{equation}
k(0) = \sqrt{2m(E_{kin}-V(0))}=\sqrt{2 m E_{kin}} \sqrt{1-\frac{V(0)}{E_{kin}}}
\simeq k \Biggl( 1 - \frac{V(0)}{2 E_{kin}} \Biggr),
\end{equation}
such that
\begin{equation}
f(0) \simeq 1 - \frac{V(0)}{k v} \simeq \frac{k(0)}{k},
\end{equation}
i.e. it is found that the probability density is enhanced by the ratio of the
central and asymptotic momenta $k(0)/k$, instead of $(k(0)/k)^2 \simeq
((E-V(0))/E)^2$ in the highly relativistic case.
One may ask how the non-relativistic and the highly relativistic regime
are connected. A classical relativistic analysis of the particle trajectories
shows that the central focusing is given by the expression
\begin{equation}
f(0)=\frac{k(0)}{k} \frac{E-V(0)}{E},
\end{equation}
which interpolates between the non-relativistic and relativistic regime and which
is given here, for the sake of brevity, as a result without proof.


\begin{thebibliography}{99}

\bibitem{Rutherford}
E. Rutherford, Philosophical Magazine {\bf{6}}, 669-688 (1911).

\bibitem{Whitney} R.R. Whitney, I. Sick, J.R. Ficenec,
R.D. Kephart, W.P. Trower, Phys. Rev. {\bf{C9}}, 2230-2235 (1974).

\bibitem{Benhar95} O. Benhar, A. Fabrocini, S. Fantoni, I. Sick,
Phys. Lett. {\bf{B343}}, 47-52 (1995).

\bibitem{Rohe}
D. Rohe et al.,
Phys. Rev. Lett. {\bf{93}}, 182501 (2004).

\bibitem{Day89}
D. B. Day, J. S. McCarthy, Z. E. Meziani, R. C. Minehart,
R. M. Sealock, S. T. Thornton, J. Jourdan, I. Sick,
B. W. Filippone, R. D. McKeown, R. G. Milner, D. H. Potterveld,
Z. Szalata,
Phys. Rev. {\bf{C40}}, 1011-1024 (1989).

\bibitem{MezianiCoulomb}
J. Morgenstern, Z.E. Meziani,
Phys. Lett. {\bf{B515}}, 269-275 (2001). 

\bibitem{Lenz} F. Lenz, Ph.D. thesis, Freiburg, Germany (1971).

\bibitem{Knoll} J. Knoll, Nucl. Phys. {\bf{A223}},
462-476 (1974).

\bibitem{Giusti} C. Giusti, F.D. Pacati, Nucl. Phys. {\bf{A473}},
717-735 (1987).

\bibitem{Rosenfelder} F. Lenz, R. Rosenfelder, Nucl. Phys.
{\bf{A176}}, 513-525 (1971).

\bibitem{Giusti88}
C. Giusti, F. D. Pacati, Nucl. Phys. {\bf{A485}}, 461-480 (1988).

\bibitem{Traini88}
M. Traini, S. Turck-Chieze, A. Zghiche,
Phys. Rev. {\bf{C38}}, 2799-2812 (1988).

\bibitem{Traini95}
M. Traini, M. Covi, Nuovo Cim. {\bf{A108}}, 723-736 (1995).

\bibitem{Rosenfelder80}
R. Rosenfelder, Annals Phys. {\bf{128}}, 188-240 (1980).

\bibitem{Kosik}
K. S. Kim, L. E. Wright, Y. Jin, D. W. Kosik,
Phys. Rev. {\bf{C54}}, 2515-2524 (1996).

\bibitem{Yennie64}
D.R. Yennie, F.L. Boos, D.G. Ravenhall,
Phys. Rev. {\bf{137}}, B882-903 (1965).

\bibitem{Sucher}
M. Levy, J. Sucher,
Phys. Rev. {\bf{186}}, 1656-1670 (1969).

\bibitem{Blankenbecler}
R.L. Sugar, R. Blankenbecler,
Phys. Rev. {\bf{183}}, 1387-1396 (1969).

\bibitem{Wallace1}
S.J. Wallace,
Annals Phys. {\bf{78}}, 190-257 (1973).

\bibitem{Wallace2}
S.J. Wallace, J.A. McNeil,
Phys. Rev. {\bf{D16}}, 3565-3580 (1977).

\bibitem{Abarbanel}
H. Abarbanel, C. Itzykson,
Phys. Rev. Lett. {\bf{23}}, 53-56 (1969).

\bibitem{Aste0}
A. Aste, K. Hencken, D. Trautmann,
Eur. Phys. J. {\bf{A21}}, 161-167 (2004).

\bibitem{Aste1}
A. Aste, K. Hencken, J. Jourdan, I. Sick, D. Trautmann,
Nucl. Phys. {\bf{A743}}, 259-282 (2004).

\bibitem{Aste2}
A. Aste, J. Jourdan,
Europhys. Lett. {\bf{67}}, 753-759 (2004). 

\bibitem{jlab}
S. Choi, J.P. Chen, Z.-E. Meziani,
{\emph{Precision measurement of longitudinal and transverse
response functions of quasi-elastic electron scattering in the
momentum transfer range $0.55$ GeV $\le$ $|\vec{q}|$ $\le$ $1.0$ GeV}},
TJNAF Proposal E01-016 (2005).

\bibitem{Traini2001}
M. Traini,
Nucl. Phys. {\bf{A694}}, 325-336 (2001).

\bibitem{Saclay}
A. Zghiche, J.F. Danel, M. Bernheim, M.K. Brussel, G.P. Capitani,
E. De Sanctis, S. Frullani, F. Garibaldi, A. Gerard,
J.M. Le Goff, A. Magnon, C. Marchand, Z.E. Meziani, J. Morgenstern,
J. Picard, D. Reffay-Pikeroen, M. Traini, S. Turck-Chieze, P. Vernin,
Nucl. Phys. {\bf{A572}}, 513-559 (1994),
Erratum ibid. {\bf{A584}}, 757 (1995).

\bibitem{Benhar94}
O. Benhar, A. Fabrocini, S. Fantoni, I. Sick,
Nucl. Phys. {\bf{A579}}, 493-517 (1994).

\bibitem{Benhar}
O. Benhar, D. Day, I. Sick, nucl-ex/0603029 (2006).

\bibitem{Udias}
J.M. Udias, P. Sarriguren, E. Moya de Guerra, E. Garrido, J.A. Caballero,
Phys. Rev. {\bf{C48}}, 2731-2739 (1993).

\bibitem{Jin}
K.S. Kim, L.E. Wright, Yanhe Jin, Phys. Rev. {\bf{C54}}, 2515-2524 (1996).

\bibitem{Aste3}
A. Aste, C. von Arx, D. Trautmann, Eur. Phys. J. {\bf{A26}}, 167-178 (2005).

\bibitem{TBR83} D. Trautmann, G. Baur, F. R\"osel,
J. Phys. B: At. Mol. Phys. {\bf{16}}, 3005-3013 (1983).

\bibitem{Pauli}
H. C. Pauli, U. Raff, Comp. Phys. Commun. {\bf{9}}, 392-407 (1975).

\bibitem{JourdanCoulomb}
J. Jourdan,
Nucl. Phys. {\bf{A603}}, 117-160 (1996).

\bibitem{Baker}
A. Baker, Phys. Rev. {\bf{D6}}, 3462-3469 (1972).


\end{thebibliography}
\end{document}